\documentclass[prx,twocolumn]{revtex4-2}
\pdfoutput=1
\usepackage{graphicx}
\usepackage{amsmath}
\usepackage{natbib}
\usepackage{epstopdf}
\usepackage{xcolor}
\newcommand{\n}{\nonumber}
\newcommand{\bn}{\begin{eqnarray}}
\newcommand{\en}{\end{eqnarray}}
\newcommand{\eml}{\end{multline}}
\newcommand{\bml}{\begin{multline}}
\newcommand{\h}{\hspace}

\newcommand{\op}[1]{\hat{#1}}

\newcommand{\pt}{\partial_x}

\begin{document}

\title {Effects of a rotating periodic lattice on coherent quantum states in a ring topology: The case of negative nonlinearity}

 \author{Jonathan Tekverk $^1$, Christopher Siebor $^1$, and Kunal K. Das$^{1,2}$}
 \affiliation{$^1$Department of Physics and Astronomy, Stony Brook University, New York 11794-3800, USA}
 \affiliation{$^2$Department of Physical Sciences, Kutztown University of Pennsylvania, Kutztown, Pennsylvania 19530, USA}
%\date{\today }
%
\begin{abstract}
We study the spectrum and stationary states in a ring-shaped lattice potential in the context of ultracold atoms with attractive interatomic interactions. We determine analytical solutions in the absence of a lattice by mapping them to those for repulsive interactions, and then we numerically follow the transformation of those solutions as the lattice is introduced and strengthened. Several features emerge that are specific to negative nonlinearity, that include: Soliton branches detaching to create new ground states; gaps opening up at the bottom of the primary spectral branch; multiple splitting and rejoining of some branches. We correlate the spectral features with the behavior of the density and phase of the corresponding eigenstates, and track them along branches and as various system parameters change. We find that the phase is sensitive to how a specific point in the spectrum is approached, particularly relevant at certain persistent gaps in the spectrum. The symmetry and stability properties are generally found to be opposite of that found for repulsive interactions.
\end{abstract}

\maketitle

\section{Introduction}

This paper is the second of a sequence of two papers that examines the stationary states of a self-interacting coherent medium in a ring-shaped lattice potential. The first paper considered  repulsive interaction \cite{Das-Huang}, and in this current one, we consider attractive interaction.  Our analyses are framed in the context of a Bose-Einstein condensate (BEC) in a toroidal trap \cite{ramanathan}, described with a mean field picture \cite{RMP-Sringari-1999} based on a nonlinear Schr\"odinger equation (NLSE).

We are motivated by a rich array of physical phenomena available when a BEC is confined within the non-trivial topology of a ring trap with an added azimuthal periodic lattice structure; which have been studied with both continuum \cite{Das-PRL-localization,Opatrny-Kolar-Das-rotation,Das-Brooks-Brattley, Aghamalyan-two-ring-lattice, Tiesinga-soliton-lattice,  Jezek-winding-number, Nigro_2018, Opatrny-Kolar-Das-LMG,Das-Christ} and discrete \cite{Doron-Cohen-1,Moreno-Bose-Hubbard,Jezek-Bose-Hubbard-ring-lattice,Maik-dipolar,Piza-ring-lattice,
Minguzzi-PRA-two-bosons,Penna, Aghamalyan-AQUID, Minguzzi-resonant-persistent,Arwas-Cohen-PRA2017,Arwas-Cohen-PRA2019} models. Toroidal traps have been utilized in several experiments to explore physics associated with the superfluidity of BEC \cite{Campbell_resistive-flow,Phillips_Campbell_superfluid_2013,Phillips_Campbell_hysteresis} but usage of an azimuthal periodic lattice is pending despite existing capabilities \cite{Padgett, Zambrini:07}.

Interactions are naturally present in ultracold atomic systems \cite{Bloch-RMP-Many-Body}, and makes the physics richer and more complex. Specifically, in the mean field limit, the setup we consider can be used to probe and simulate nonlinear dynamics in a lattice system which is closed, finite and unbounded. A necessary preliminary to exploring the dynamics is a thorough understanding of the full landscape of stationary solutions. We do that here for attractive interaction to complement our prior analysis with repulsive interaction \cite{Das-Huang}, to present a comprehensive picture.

In lattices with trivial topology, experiments \cite{Oberthaler-gap-soliton,Oberthaler-selftrapped} have explored interacting BEC, complemented by theoretical studies \cite{RMP-solitons,Bronski, Wu_Niu,Konotop_Salerno,Pethick_Smith_PRA,smith03, smith04,Holland-Kronig-Penney,holland05}.  However, in the context of ring lattices, apart from our recent work \cite{Das-Huang}, there has only been one other detailed study \cite{Guilleumas-nonlinear_ring}, but none for attractive interactions. Our approach here will be built on the analytical solutions that exist for such nonlinear systems in the absence of a lattice \cite{carr00-a,carr00-b} and determine the effects of introducing a lattice of increasing strength. This will draw on a convenient mapping between solutions for positive and negative nonlinearities that one of us recently demonstrated for the relevant NLSE \cite{Brattley-Das-PRA}.

Our results will have relevance in a broader context since the nonlinear Schr\"odinger equation we use is also relevant in the field of nonlinear and fiber optics \cite{NLSE-Chiao-Townes, Boyd, G.P.Agarwal}. Thus, it is well established that attractive interaction leads to significantly different behavior, for example, in three dimensions sufficiently strong interactions and large number of particles can lead to wave-function collapse \cite{Sackett-collapse}, but though not necessarily in one dimension (1D) \cite{Carr-1D-BEC}. There have been no comprehensive study of how attractive interactions in an NLSE impact the states in the presence of a lattice in a ring configuration. As we will show in the paper, there are indeed dramatically different behaviors manifest when the interaction becomes attractive that we did not observe in our prior study of the same configuration with repulsive nonlinearity.

In Sec.~II, we summarize our physical model; and then we determine the analytical solutions in the absence of a lattice in Sec.~III, focussing on the essential features that distinguish having attractive interactions instead of  repulsive. We discuss the effects of introducing the lattice, on the spectrum in Sec.~IV; and proceed to examine the impact on the corresponding stationary states, first on the density in Sec.~V and then on the phase in Sec.~VI.  The interplay between varying the strength of the lattice and that of the  interaction-induced nonlinearity is explored in Sec.~VII as regards the effects on the spectrum. In Sec.~VIII, we focus on several anomalous spectral features we observe. Section IX presents a stability analysis for the various stationary states. We conclude with a summary of our results and outlook in Sec.~X.

\begin{figure}[t]
\centering
%\vspace{-0.2\linewidth}
\includegraphics[width=\columnwidth]{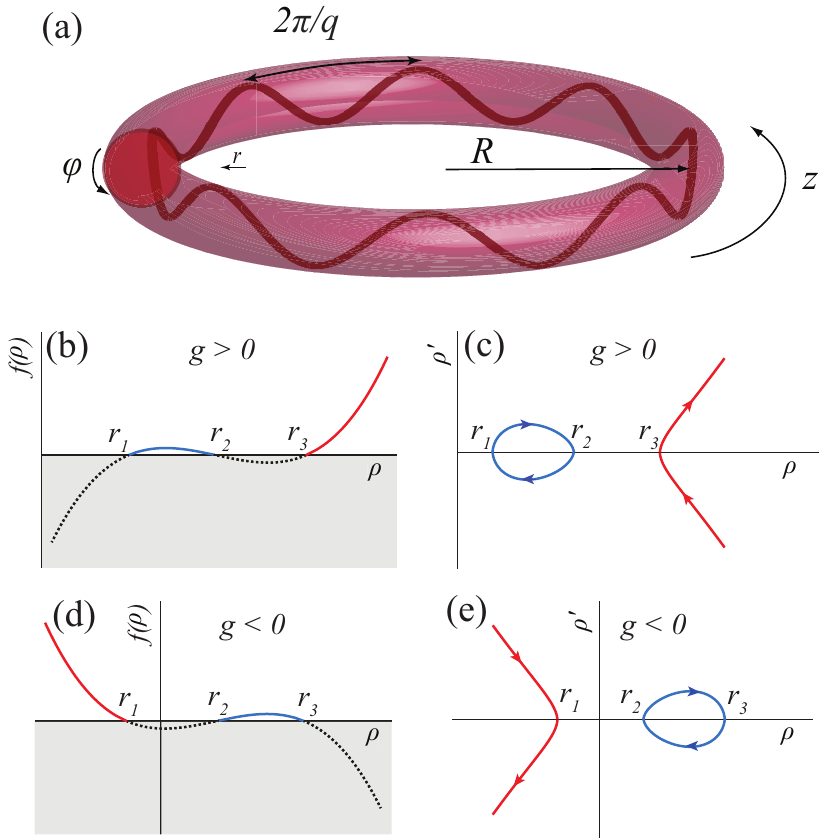}
\caption{(a) Atoms are trapped in a toroidal trap with an azimuthal lattice potential of period  $2\pi/q$, its variation of depth shown schematically as a thick sinusoidal line. The torus is taken as a wrapped cylinder with our choice of co-ordinates ${\bf r}=(z,r,\varphi)$ shown, assuming the major radius to be much larger than the minor radius, $R\gg r$. Features of phase space curves are compared for (b,c) positive nonlinearity $g>0$ with those for (d,e) negative nonlinearity $g<0$. (b,d) Schematics of  the cubic function $f$ that sets the density variation, shown for all three real roots $\{r_i\}$; the dotted parts lie in the shaded forbidden region. (c,e) Corresponding phase space plot of $\rho'=\pm\sqrt{f}$ versus $\rho$ has a wing-loop structure; the orientation of the wing depends on the sign of the nonlinearity $g$. }
\label{Fig1_functionshape}
\end{figure}

\section{Physical Model}

The details of our physical model are presented in our prior paper focussed on repulsive interactions \cite{Das-Huang}; here we will only recapitulate the salient points. Consider a BEC in a toroidal trap with its minor radius $r$ much smaller than its major radius $R$ so that the system is treated as a cylinder ${\bf r}=(z,r,\varphi)$ with periodic boundary condition on $z$.  Strong  confinement along $(r,\varphi)$ justifies integrating them  out for an effective 1D Hamiltonian
\begin{eqnarray}\label{QF-Hamiltonian}
\op{H}(t)&=&\int_0^{2\pi R}{\rm d}z\op{\Psi}^\dagger(z,t)\times\left[-\frac{\hbar^2}{2m}\partial_z^2+V(z,t)\right.\\
&&\left.+\frac{g_{3D}N}{4\pi l^2} \op{\Psi}^\dagger(z,t)\op{\Psi}(z,t)\right]\op{\Psi}(z,t).\n
\end{eqnarray}
where  $g_{3D}=4\pi\hbar^2a/m$ is the interaction strength, $a$ is the $s$-wave scattering length,  $m$ is the mass of individual atoms, $N$ is the total number of atoms, and $l=\sqrt{\hbar/m\omega_T}$ is the harmonic oscillator length for the transverse confinement along the minor radius. Taking the major radius $R$ as the length unit, the azimuthal distance is rescaled $z/R=\theta\in [0,2\pi)$;  the lowest energy scale in the ring $E_R=\frac{\hbar ^2}{mR^2}$ is set as the energy unit; and the corresponding frequency $\omega_R=\frac{\hbar}{mR^2}$ is set as the unit for frequency as well as for angular velocity, and sets the time scale $\tau=\omega_R^{-1}$. The effective nonlinear constant in 1D then has the form $g=2a\omega_TN$.  Using these units leads to the equation of motion which, in the mean field limit $\langle\hat{\Psi}\rangle=\psi$, is a nonlinear Schr\"odinger equation:
\begin{equation}
\label{eq1.1-9}
\left[{\textstyle\frac{1}{2}}(i\partial_\theta+\Omega)^2+V_0\sin^2({\textstyle\frac{1}{2}}q\theta)+g|\psi|^2\right]\psi=i\partial_t\psi
\end{equation}
with normalization $\int_0^{2\pi}d\theta \left|\psi(\theta,t)\right|^2 =1$. The lattice $V(z,t)=V_0\sin^2\left(\frac{q}{2}(z/R-\Omega t)\right)$ can rotate with angular velocity $\Omega$ with respect to the laboratory frame, and in Eq.~(\ref{eq1.1-9}) we transformed to the frame rotating with the lattice, so the Hamiltonian itself is time-independent. The stationary solutions $\varphi(\theta)=\psi(\theta,t)e^{i\mu t}$ satisfy the time-independent version of Eq.~(\ref{eq1.1-9}) with $i\partial_t\rightarrow \mu$ where the eigenvalues $\mu$ defines the chemical potential.

The stationary solutions can be determined in the hydrodynamic picture, expressing $\varphi(\theta)=\sqrt{\rho(\theta)}e^{i\phi(\theta)}$, leading to coupled equations for the density $\rho$ and the phase $\phi$; the latter can be formally integrated:
\bn
\label{eq1.2-2}
{\textstyle\frac{1}{8}}(\pt\rho)^2\h{-2mm}-{\textstyle\frac{1}{4}}\rho\pt^2\rho+{\textstyle\frac{1}{2}}\alpha^2\h{-1mm}+V_0\sin^2({\textstyle\frac{1}{2}}q\theta)\rho^2\h{-1.5mm}+g\rho^3\h{-1.5mm}-\mu\rho^2=0
\n\\
\Delta \phi (\theta)=\phi (\theta)-\phi (0) =\Omega\theta+\int_{0}^{\theta}\frac{\alpha}{\rho(\theta')}d\theta'.\h{1cm}
\en
The integral of motion is associated with the current density $J=N\alpha$ and the superfluid velocity $v=\alpha/\rho(\theta)$. The ring topology imposes periodic boundary conditions
\bn
\label{eq1.3-2}
\rho(0)=\rho(2\pi) &&
\rho'(0)=\rho'(2\pi) \n\\
\Delta \phi (2\pi)
=2\pi n,&&
\delta\phi\equiv\delta\phi(2\pi)= \Delta\phi(2\pi)-2\pi\Omega,
\en
with the integer $n$ being the winding number.  We will plot the chemical potential $\mu$ as a function of the bare phase change $\delta\phi$ acquired around the ring neglecting rotation.  The phase boundary condition for a finite size ring picks out only discrete points on the continuum spectrum as physically relevant but, as can be seen above, the entire spectrum can be accessed via rotation \cite{Das-Huang}.

\section{Exact solutions without lattice}
\label{Exact_solutions}
In the absence of a lattice, $V_0=0$, analytical solutions can be found for Eq.~(\ref{eq1.2-2})  \cite{Das-Huang, Brattley-Das-PRA, carr00-a,carr00-b}, and they will provide the framework for our description once the lattice is introduced. We will show that the solutions for $g<0$ can be related intuitively to those for $g>0$ we presented earlier \cite{Das-Huang}. Integration of Eq.~(\ref{eq1.2-2}) yields
\bn
\label{eq.hd.3}
\pt\rho=\pm\sqrt{f\left(\rho\right)}; \h{3mm}
f\left(\rho\right)=4g\rho^3-8\mu\rho^2+8\beta\rho-4\alpha^2
\en
with integral of motion $\beta$. All the stationary states of the system can be expressed in terms of the three roots of this cubic function $f(\rho)$ \cite{Brattley-Das-PRA}. The function is compared for $g>0$ and $g<0$ in Fig.~\ref{Fig1_functionshape}(b,d). With the forbidden negative density regimes of the function (shown in dotted lines) left out, the phase space plots of $\rho'=\pm\sqrt{f(\rho)}$ versus $\rho$ in Fig.~\ref{Fig1_functionshape} (c,d) generally form a ``loop-wing" structure, with a closed loop that corresponds to oscillating solutions and an open wing shape tied to solutions that diverge. Special cases occur when some of the roots are degenerate or complex \cite{Brattley-Das-PRA}.  In terms of the roots, the coefficients are
\bn\label{parameters}\textstyle
 \mu=\frac{g}{2}(r_1+r_2+r_3),\h{5mm}\alpha^2&=& gr_1 r_2 r_3\n\\
\beta=\frac{g}{2}(r_1 r_2+r_1 r_3+r_2 r_3).\en
For real roots, without loss of generality, we assume ascending order of the roots $r_1\leq r_2\leq r_3$.

Since $\alpha^2\geq 0$, one root has to be negative or zero when $g<0$; but two and three negative roots are not allowed; and neither are complex roots since a conjugate pair would have a positive product and the whole phase space curve would be in the unphysical negative density regime. All the roots therefore have to be real with $r_1\leq 0$, and $r_2,r_3 \geq 0$; and the wing intersects with the $\rho=0$ axis at $r_1$. The wing, being open on the left, lies entirely in the $\rho\leq 0$ regime.  Therefore solutions on the wing are forbidden, and only those on the loop are allowed.  Furthermore, in a ring with uniform potential, solutions that grow or decay will not satisfy the periodic boundary conditions. So, only oscillatory solutions are possible, with planes waves being a limiting case.

\begin{figure*}[t]
    \centering
    \includegraphics[width=1\linewidth]{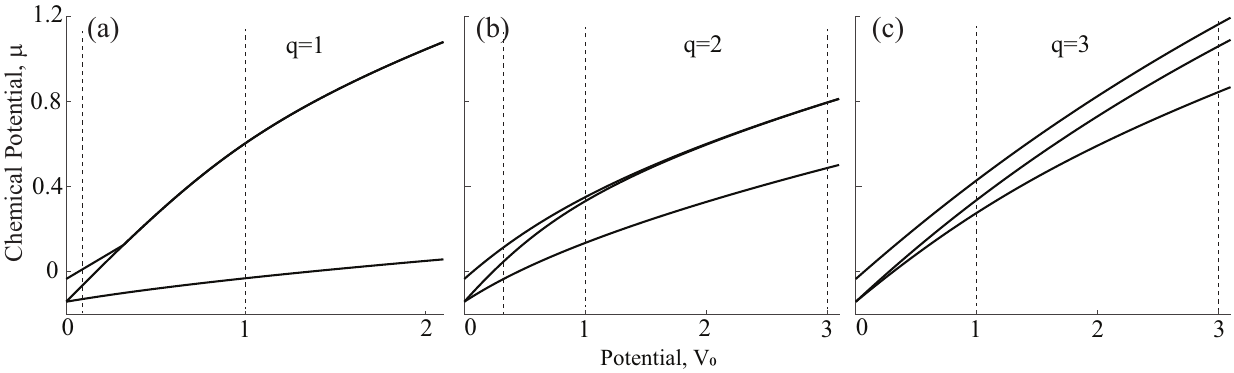}
    \caption{ The chemical potential $\mu$ plotted as a function of the lattice depth $V_0$ for the number of lattice periods $q=1,2,$ and $3$ in panels (a,b,c). The nonlinear strength is fixed at $g=-1$ and the bare phase  accumulated around the ring $\delta\phi=0.5$ correspond to the vertical dashed lines in Fig.~\ref{Fig3_spectra_vs_phase}.  The vertical dashed lines that appear here in turn indicate the values of $V_0$ that correspond to the plots of $\mu$ versus $\delta\phi$ that appear in Fig.~\ref{Fig3_spectra_vs_phase}.}
    \label{Fig2_spectra_vs_V}
\end{figure*}

The general shape of the phase space curves for $g>0$ and $g<0$ are mirror reflections across some symmetry axis, switching the roles of $r_1$ and $r_3$.  As a result, we find that solutions for $g<0$ can obtained from corresponding solutions for $g>0$ by simply switching $r_1\leftrightarrow r_3$. We proved this rigorously using the properties of Jacobi elliptic functions in Ref.~\cite{Brattley-Das-PRA}.  This insight allows us to summarize the solutions for the attractive interactions as direct counterparts of the solutions we found for the repulsive interactions in our prior work \cite{Das-Huang}, with the density and phase now given by
\bn\label{general_solution}
\rho(\theta)&=& r_3+(r_2-r_3){\rm sn} ^2(\textstyle{\sqrt{g(r_1-r_3)}}\ \theta,m') \n\\
\delta\phi(\theta)&=&\frac{\alpha}{r_3\sqrt{g(r_1-r_3)}}\Pi(1-\frac{r_2}{r_3},\varphi,m')
\en
where $\Pi$ is an incomplete elliptic integral of the third kind \cite{byrd13book} and $\varphi=\sin^{-1}[{\rm sn}(\sqrt{g(r_1-r_3)}\ \theta,m')]$. The parameter $m'$ can be obtained from its counterpart, $m$ for $g>0$ by switching  $r_1\leftrightarrow r_3$ as well
\bn\label{parameters}\textstyle
m'=1-m&=&\frac{r_2-r_3}{r_1-r_3}.\en
The density oscillates with period $\theta=K(m')/\sqrt{g(r_1-r_3)}$. To satisfy the density boundary condition, the \emph{complete} phase space loop has to be traversed by an integer number of turns $j$, leading to the condition for a complete circuit of the ring,
\begin{equation}
\label{eq.hd.8}
jK(m')=\pi\sqrt{g(r_1-r_3)}.
\end{equation}
The bare phase in a circuit of the ring and the normalization of the density constrain the solutions
\bn \label{bare_phase}\delta\phi=\delta\phi(2\pi)=\frac{2\pi \alpha}{K(m') r_3}\Pi(1-\frac{r_2}{r_3},m')\h{1cm}\\
\textstyle \int_{0}^{2\pi}\h{-2mm}\rho d\theta=2\pi r_3+2\pi(r_1-r_3)[1-E(m')/K(m')]=1. \n\en
Here, $K(m')$ and $E(m')$ are complete elliptic integrals of the first and the second kinds, respectively.

The definition of $m'$ in Eq.~(\ref{parameters}) along with Eqs.~(\ref{eq.hd.8}) and (\ref{bare_phase}) can be used to express the roots in terms of the complete elliptic integrals of the first and second kinds, exact counterparts of those we derived for $g>0$ after applying the mapping we mention above. When $j$ and $g$ are specified, these are completely determined by the value of $m'$, which can be determined by imposing the phase boundary condition, yielding the complete solution.

The spectrum plotted as a function of the bare phase comprises of a parabolic dispersion curve and a sequence of swallowtail branches \cite{Wu_Niu-landau-zener} that mark solitonic solutions.  As with positive nonlinearity, we find three distinct types of the solutions: (i) Plane waves corresponding to the spectral values on the parabolic dispersion curve; (ii) Density modulations with nodes that correspond to the tips of the swallowtails and (iii) Nodeless density modulations that mark the rest of the swallowtail branches.

\begin{figure*}[t]
    \centering
    \includegraphics[width=1\linewidth]{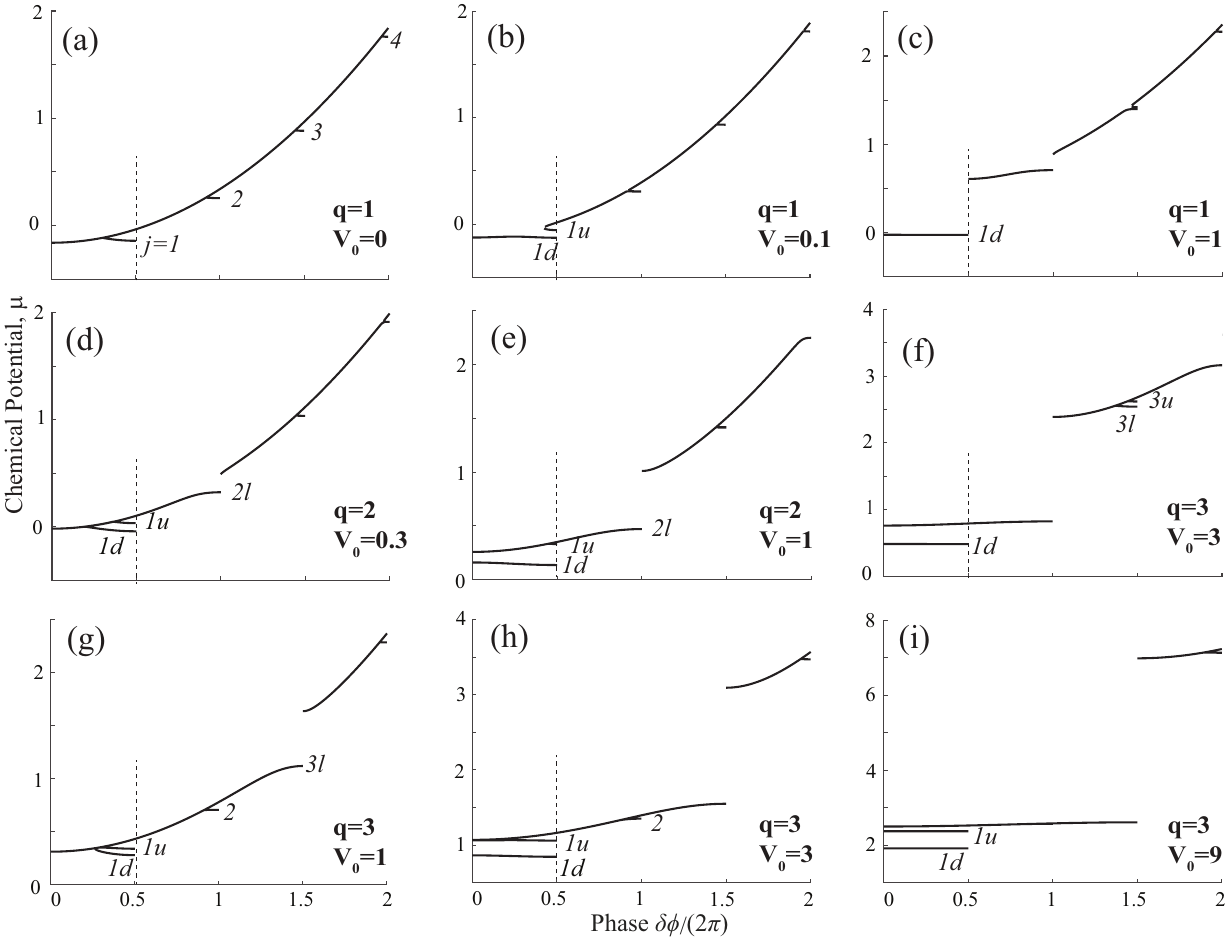}
    \caption{ The chemical potential $\mu$ plotted as a function of the bare phase $\delta\phi$ for three different lattice periods $q=1,2,$ and $3$, one in each row.  The nonlinear strength is fixed at $g=-1$. The three panels in each row are for different values of the lattice depth $V_0$ corresponding to the vertical dashed lines in Fig.~\ref{Fig2_spectra_vs_V}. The vertical dashed lines here indicate the bare phase values $\delta\phi$ that correspond to the plots in Fig.~\ref{Fig2_spectra_vs_V}.}
    \label{Fig3_spectra_vs_phase}
\end{figure*}

We discussed the features of these solutions for $g>0$ in some detail \cite{Das-Huang}, here we will discuss the differences that emerge for $g<0$.  For plane wave solutions, $g<0$ shifts the parabola down $\mu=\frac{n^2}{2}+\frac{g}{2\pi}$, which translates to a downward shift for the entire spectrum; so the ground state has negative energy. The bare phase in Eq.~(\ref{bare_phase}) as function of $m'$ is monontically increasing for $g<0$ and so, for the swallowtail branches, the phase lies in the range $[\sqrt{j^2\pi^2+2\pi g},j\pi]$ corresponding to $m'\in [0,m'_c]$. The $m=0$ value corresponds to the lower limit of the  bare phase for negative $g$ and corresponds to the end attached to the parabolic curve, where it tends to a plane wave. The value $m_c'$ correspond to bare phase multiples of $\pi$, the tips of the swallowtails which mark the solutions with nodes. Since the upper value of the bare phase marks the tip, the swallowtails are now on the outside of the parabola as we will see in Fig.~\ref{Fig3_spectra_vs_phase} discussed in the next section. This is in contrast to those for positive nonlinearity, where the swallowtails extend on the inside concave side of the parabola \cite{Das-Huang}. This is the most striking difference in the spectrum between $g>0$ and $g<0$, in the absence of lattice.

There is another distinguishing feature for these swallowtails, that remains significant even when the lattice is present: Clearly for $g<-j^2\pi/2$, the range for the bare phase becomes $[0,j\pi]$ which, as we will see, marks a transition point where the swallowtail branch of index $j$ detaches from the parabola and slides below the lowest energy possible for the plane wave solutions associated with the parabola; and the ground state then corresponds to a swallowtail. This behavior is not seen for repulsive interactions when $g>0$.

The solutions with nodes cannot carry current, and therefore can be made real-valued everywhere, and they can be readily obtained from the general solution in Eq.~(\ref{general_solution}) setting $r_2=0$
\begin{equation}
\label{eq2.1-1}
\begin{array}{c}
\psi(\theta)=\sqrt{\frac{m'}{2\pi\left(m'+E/K-1\right)}}{\rm cn}\left(\frac{jK\theta}{\pi}| m'\right),\
\mu=\frac{(1+m')j^2K^2}{2\pi^2}.
\end{array}
\end{equation}
This differs in having an elliptic ${\rm cn}$ rather than ${\rm sn}$ that is present for $g>0$, and in that the elliptic parameter is $m'$ instead of $m$.

\section{Spectrum with Lattice}

In the presence of attractive interatomic interactions, $g<0$, even with no lattice, there can be a multitude of solutions. Keeping the interaction strength fixed at $g=-1$ and the bare phase gathered around the ring fixed at $\delta\phi=0.5$, we track the progression of the chemical potential in Fig.~\ref{Fig2_spectra_vs_V} as the lattice potential is turned on, for three different lattice periods, $q=1,2,3$. The qualitative behavior here resembles that for $g>0$: Some of the branches split and then some of them rejoin again.

Significant differences in behavior emerge when we plot the chemical potential as a function of the bare phase in Fig.~\ref{Fig3_spectra_vs_phase}. In these and all other similar spectrum plots in the paper, we reiterate that for a fixed value of the angular velocity (including $\Omega=0$) only specific points on each trace are allowed that meet the phase boundary condition in Eq.~(\ref{eq1.3-2}); those points lie on equally spaced vertical slices in the spectral plots such as in Fig.~\ref{Fig3_spectra_vs_phase} and any angular velocity $\Omega$ shifts those lines continuously \cite{Das-Huang} allowing access to the entire spectrum.  Even when there is no lattice, the obvious difference in changing the sign of the nonlinear is that for $g<0$ the soliton branches are on the outside of the main parabola which marks plane wave solutions; for $g>0$ they are on the inside. As the lattice is turned on, this difference leads to several interesting features that differentiate negative nonlinearity from positive. What remains similar is that the lattice splits the soliton branches, but with gaps progressively smaller for higher $j$, requiring stronger lattices to create noticeable gaps. Gaps open up in the parabola as well at the soliton branches  with index $j$ that are commensurate, as in exact multiples of  the lattice periodicity, $q$.

When $q=1$, there is only one lattice period around the ring, and every soliton index $j$ is commensurate. So in Fig.~\ref{Fig3_spectra_vs_phase}(a-c), as the lattice is introduced, band gaps open up on the parabola at every soliton branch.  For $q=2$ in Fig.~\ref{Fig3_spectra_vs_phase}(d-f) and $q=3$ in Fig.~\ref{Fig3_spectra_vs_phase}(g-i), distinctions emerge between inter-band and intra-band soliton branches, corresponding to the soliton index being commensurate and incommensurate respectively. As lattice strength increases, the splits appear in all the soliton branches,though progressively less conspicuous for higher ones. For commensurate branches, the split soliton branches blend with the main branch segments above and below the gap to form hook-like structures which, for negative nonlinearity, lie on the \emph{outer} convex side of the parabola, as shown by Fig.~\ref{Fig3_spectra_vs_phase}(b-c). With increasing lattice depth, the spectral lines  flatten out, causing the commensurate soliton branches to become visibly indistinguishable from the main branches at band gaps.

The intraband branches highlight a different behavior: For positive $g$, with increasing  lattice strength those branches slide upward towards higher chemical potential.  With negative $g$, the opposite happens and they slide downward along the parabola. In the case of the lowest band, at sufficiently high lattice strength, those spectral lines eventually `fall off' the main branch and connect directly to the vertical axis, creating a new ground state in the process. This behavior can be seen for both $q=2$ and $q=3$ in Fig.~\ref{Fig3_spectra_vs_phase}, with the $j=1$ branch detaching and falling below the main branch in both cases.

We noted in the previous section that this behavior occurs even without a lattice, for any branch when $g\leq -j^2\pi/2$. Here we see  that even at values of $g$ that does not meet that criterion, increasing the lattice strength induces the same qualitative behavior. The strength of the nonlinearity $g$ determines the number of intra-band soliton branches that detach even with no lattice.

This underscores the intricate interplay of the lattice and the nonlinearity that defines the behavior of this system.  That interplay is responsible for other interesting variations of this behavior, some seen in Fig.~\ref{Fig3_spectra_vs_phase}: In the case of $q=2$, the $j=1$ branch splits. Labelling the upper and lower branches $u$ and $d$ for `up' and `down', only the lower branch $1d$ detaches from the main branch, while the upper one $1u$ shortens and blends with the main branch as the branches flatten with increasing lattice depth. On the other hand, for $q=3$, both $1u$ and $1d$ detach and remain so as the lattice depth increases. However, we do see the $j=2$ branch vanish into the main branch. Although the lattice splits the soliton branches, the effect is not quite uniform. The commensurate branches split significantly as they coincide with the band gaps opening up.  The intraband solitons splittings are smaller.  Some branches are surprisingly resistant to any noticeable splitting; for example the $j=2$ branch for $q=3$ in Fig.~\ref{Fig3_spectra_vs_phase}(g-i) simply disappears at sufficiently deep lattice.

\begin{figure*}[t]
    \centering
    \includegraphics[width=1\linewidth]{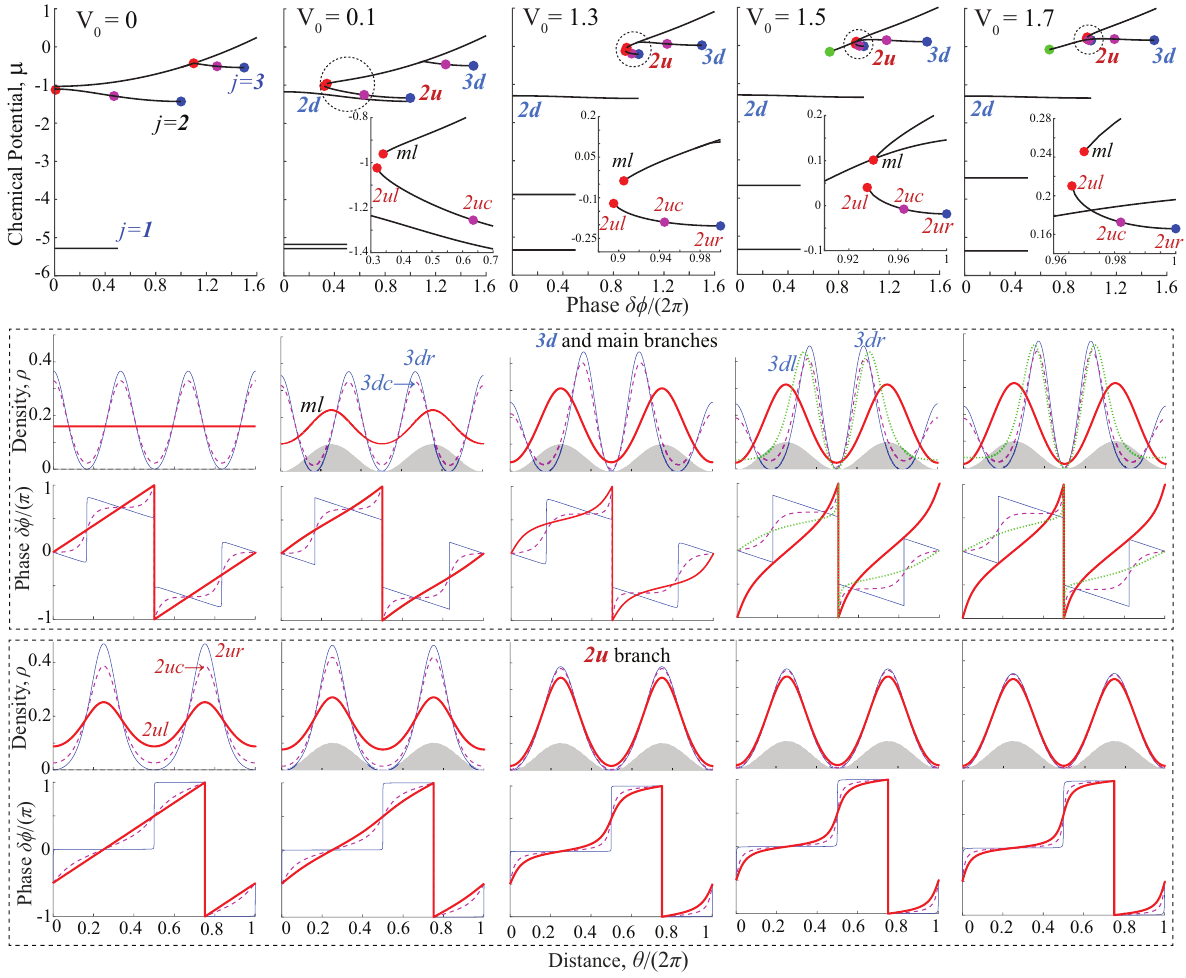}
    \caption{ The top row plots the chemical potential $\mu$ as a function of the phase $\delta\phi$, for lattice period $q=2$, and shows the progression as the lattice strength  is increased from $V_0=0$ to $V_0=1.7$, keeping the nonlinear strength fixed at $g=-6.5$. The insets show details of the upper splitting of $j=2$, labelled $2u$, and the unsplit $j=3$ branch.  The solid circles mark values for which the states are plotted  in the rows below, with labels as described in Eq.~(\ref{name-convention}). The plots of the density and phases of the state are lined up with the corresponding spectral plots in the top row, with the middle rows for soliton branch $3d$ and one point from the main branch, and the bottom rows for the soliton branch $2u$. Labels on the density traces mark which point they correspond to in the spectrum, the phase plots have matching traces. The lattice is indicated in filled gray.}
    \label{Fig4_q2_states}
\end{figure*}

\section{Effect of Lattice on Density}\label{Sec:Lattice-density}

In order to get a better insight into the correlation of the various spectral features and ultimately the behavior of the system, we have to examine the associated eigenstates. We do so in the context of  $q=2$ and $q=3$, fixing the nonlinear strength and then switching on and progressively increasing the lattice depth.

As the spectrum evolves with the introduction of the lattice, it becomes harder to classify branches as being on the main branch or a specific soliton branch indexed by $j$ as the branches blend, split and merge.  Certain features of the eigenstates help identify and classify them: (i) In the presence of a lattice, the main branch solutions transform from plane waves to Bloch waves with the same period as the lattice. (ii) Soliton branches can be identified by the number of density peaks that should equal the $j$ index. (iii) When soliton branches split, the eigenstates corresponding to the  upper and lower branches display different symmetry properties relative to the lattice.

The symmetry properties were discussed in detail in the case of positive nonlinearity \cite{Das-Huang}, here we note the differences: For $g<0$, for each splitting, the energetically lower branches  have at least one density maximum line up with a lattice minimum, while the upper branches \emph{never} have any state maximum line up with a lattice minimum. This is in contrast for $g>0$ wherein for a lower branch, at least one of the density minimum lines up with a lattice maximum; and for the upper branch, lattice maxima and state maxima never coincide.

\begin{figure*}[t]
    \centering
    \includegraphics[width=\linewidth]{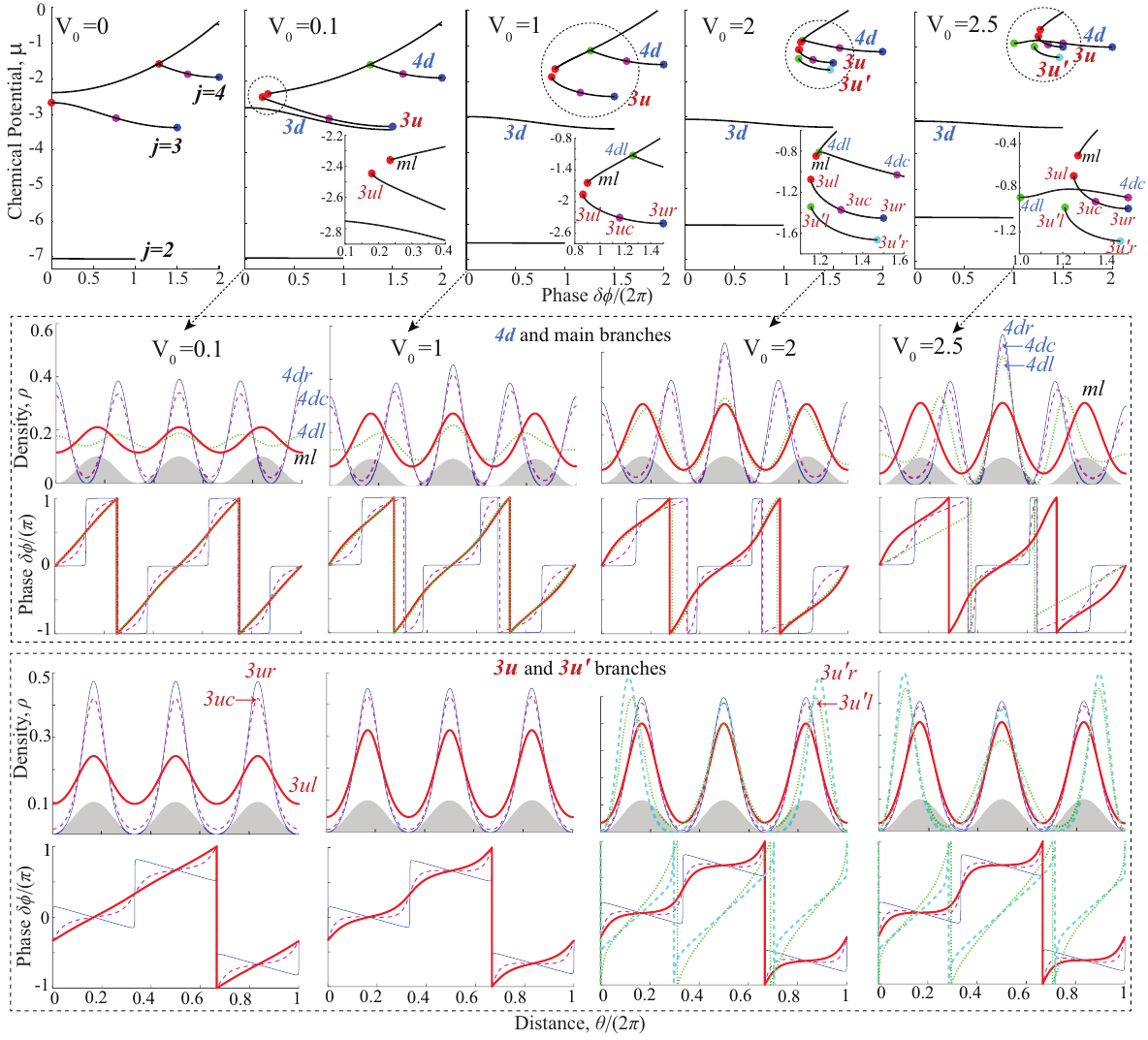}
    \caption{  The top row plots the chemical potential $\mu$ as a function of the phase $\delta\phi$, for lattice period $q=3$, and shows the progression as the lattice strength  is increased from $V_0=0$ to $V_0=2.5$, keeping the nonlinear strength fixed at $g=-15$. The inset shows details of the upper split of $j=3$, labelled $3u$ and the unsplit $j=4$ branches. The solid circles mark values for which the states are plotted  in the rows below, with labels as described in Eq.~(\ref{name-convention}). The density and phase are not plotted for $V_0=0$, so the lower four rows line up and correspond to only the non-zero potentials in the top row, as indicted by arrows. The middle rows are for soliton branch $4$ and a point from the main branch, and the bottom rows are for the soliton branch $3u$ with the last two columns including states from the newly emergent $3u'$ branch as well. Labels on the density traces mark which point they correspond to in the spectrum, the phase plots have matching traces. The lattice is indicated in filled gray.}
    \label{Fig5_q3_states}
\end{figure*}

The eigenstates as plotted in Fig.~\ref{Fig4_q2_states} and~\ref{Fig5_q3_states} provide more insights. Densities and phases of the states are plotted in vertical alignment with the corresponding spectra in the top row. On the spectra, solid dots mark the points that correspond to states plotted below, and labelled by:
\bn [j\mbox{-}index]\ [sub\mbox{-}branch]\ [location],\h{5mm} j=1,2,\cdots\n\\
sub\mbox{-}branch= \{m,u/u',d/d'\}, \h{5mm} location=\{l,c,r\}\label{name-convention}\en
with $j$ being the swallowtail branch index based on the analytical solutions with no lattice; the \emph{sub-branch} labels are $m$ for main branch, $u,d$ for up/down marking upper/lower in a split and if there are further splittings of those branches, we label them $u',d'$; and the \emph{location} labels $l,c,r$ indicate left, somewhere near the center, and right end of each branch.

\subsection{q=2}

We examine the case of $q=2$ lattice periodicity in Fig.~\ref{Fig4_q2_states} at fixed $g=-6.5$. This nonlinearity is sufficiently strong to have the $j=1$ and $j=2$ soliton branches to be detached and below the main branch even with no lattice. The lattice splits them both, with the gaps getting wider with increasing lattice depth. But, the $j=2$ branch behaves differently: The upper branch, $2u$ and the main branch both detach from the vertical axis; and their left tips approach each other, tending to form a hook-like structure. However, the left tips never actually meet in our simulations, always leaving a gap, as shown in the insets in the top row of Fig.~\ref{Fig4_q2_states}. As the lattice is strengthened, the $2u$ branch shrinks, eventually disappearing (not shown). The general behavior resembles the opening of a band gap at commensurate soliton branches, but with some differences.

We observe another interesting feature with the $j=3$ branch. We do not see any visible splitting when the lattice is introduced but, due to the observed symmetry features, we label it $3d$. Increasing lattice strength causes it to slide downwards along the main branch, until it does something surprising: It extends out beyond the now much contracted main branch, as shown for $V_0=1.5$ in Fig.~\ref{Fig4_q2_states}. At even stronger lattice, the branch extends out farther and slides farther down to intersect with the shrunken $2u$ branch. An interesting characteristic of these branches is that they have termination points both on the left and the right; the left termination points are atypically not at multiples of $\pi$ as we can see in Fig.~\ref{Fig4_q2_states} while the right termination points are.

On each soliton branch, the number of density peaks equals its $j$ index. The main branch densities start off as plane waves but, with the introduction of the lattice, they transform to uniformly modulated Bloch waves similar to states on the commensurate $j=2$ branch. Specifically, as the left tips, labelled $2ul$ and $ml$, of the detached $2u$ and the main $m$ branches approach each other, the corresponding densities shown in thick red lines, converge to the same form. They appear to be mutual continuations, although we could never get the gap to close. At stronger lattice, the main branch continues to shrink, approaching termination at $\theta=2\pi$ on the left, for a solution with two nodes; and the densities all along the shrinking $2u$ branch also tend towards that two node solution.

\begin{figure}[t]
\includegraphics[width=\columnwidth]{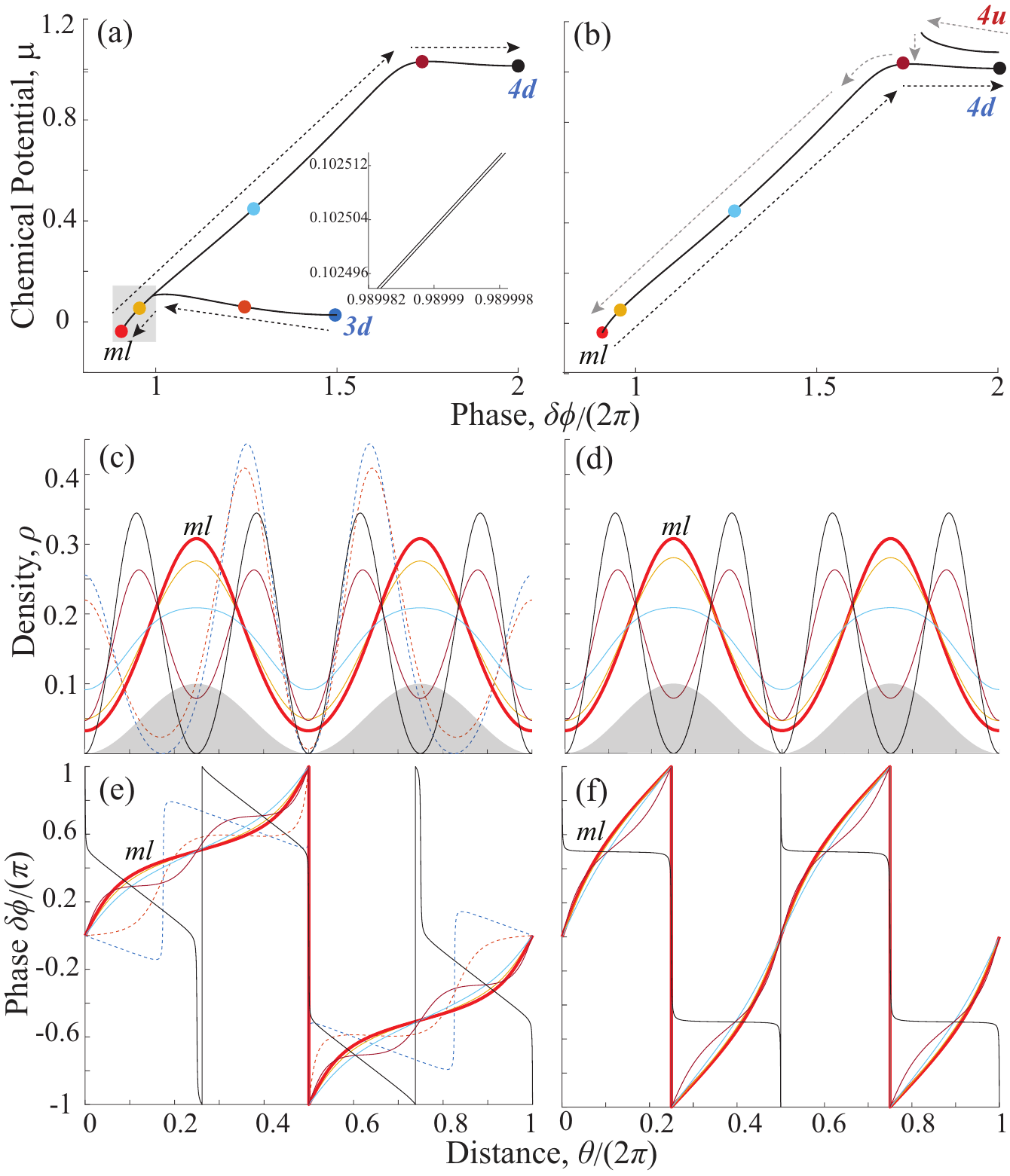}
\caption{The figure illustrates the dependence of the phase on how a particular spectral value is approached. (a,b) The chemical potential is plotted versus the phase for the main branch and $3d$ soliton branch, for $q=2$ and $V_0=1.3$, identically to the center top row panel in Fig.~\ref{Fig4_q2_states}: In panel (a) the left eigenstates are computed starting at the tip of the $3d$ branch as indicated by arrows, while in panel (b) they are computed starting at the tip of $4u$ branch. The inset in (a) shows that the bottom part of the main branch actually comprises of closely spaced double lines. (c,d) The densities are identical for corresponding points in the two different cases. (e,f) The phase however clearly behaves differently in the two cases. Two additional traces appear on the left plots (c,e) corresponding to the points on the $3d$ soliton branch.}
    \label{Fig6_phase_var}
\end{figure}

The densities on the $j=3$ soliton branch start off as Bloch waves with three peaks in the absence of a lattice, but due to it being incommensurate, its starts to get more localized with the introduction of the lattice \cite{Das-Huang}. As the lattice deepens, the right tip labelled $3dr$ continues to terminate at $\delta\phi=3\pi$ and have three nodes as it did without a lattice. As the branch shoots out and detaches from the main branch, the emergent left tip, labelled $3dl$ and the corresponding density shown in dotted green line, displays qualitatively different behavior with only two density minima.  Density at an intermediate point, $3dc$ shown in dashed magenta, indicates a morphing from three density minima at the right tip branch to only two at the left. But, the left tip does not feature nodes as it does not terminate at a phase multiple of $\pi$.

\subsection{q=3}

We next examine the case of $q=3$ lattice periodicity in Fig.~\ref{Fig5_q3_states} at fixed $g=-15$. We find counterparts of most of the features we noted above for $q=2$, adjusting for the change of the lattice period. However, since the lowest commensurate branch has a higher index $j=3$, we need a larger magnitude nonlinearity to access them. Those shared features includes the detachment of both the main branch and the $j=3u$ branch from the vertical axis, the subsequent formation of a persistent gap, as well as, the extension of the $j=4d$ branch beyond the tip of the shrunken main branch. However, there is a new feature here not seen with $q=2$:  At higher potential strength, the upper split of the $j=3$ branch splits further to create an additional branch we label $3u'$, with distinctive features that we describe below.

The $j=4$ branch does not visibly split, or at least our simulations did not converge to two separate branches. However, based on the symmetry feature that one density maximum coincides with a lattice minimum, if there is a split we have converged on the lower branch and so we label it $4d$. With increasing lattice depth, this branch behaves exactly like the $3d$ branch for $q=2$, with the density starting off as a uniform Bloch wave and getting progressively localized. Then, as it extends beyond the main branch, there is gradual morphing of the density from having $4$ nodes at the right tip to having only $2$ at the left tip, for respective bare phases of $4\pi$ and $2\pi$.

As the left tips of the detached main branch and the $3u$ branches approach each other, the gap never closes, but the densities at the their tips, labelled $ml$ and $3ul$ and plotted in thick red lines, tend to almost identical Bloch waves with three maxima, analogous to their $q=2$ counterpart in Fig.~\ref{Fig4_q2_states}.  At $V_0=2$ and above, the $3u$ branch splits further, creating the new branch $3u'$ with a density profile distinctly different from that of a Bloch wave. The densities at its left and right tips, plotted in dotted green and dashed cyan, display localization but still with three density peaks; the symmetry properties are that of an upper branch with none of the peaks lining up with a lattice minimum. The tips do not reach multiples of $\pi$, and hence they do not have nodes in their densities. They just get very close to having nodes, particularly at the right tip when the lattice depth is $V_0=2.5$ in Fig.~\ref{Fig5_q3_states}.

\section{Sensitivity of Phase}
In the previous section we observed that for both $q=2$ and $q=3$, the upper of the respectively commensurate $j=2$ and $j=3$ branches detach from the vertical axis, with the left tip tending towards that of the also detached main branch. The tips never meet even as the lattice depth increases; yet the densities seem mutual continuations, suggestive of a continuous morphing. This raises the question whether the gap is real, or just an artifact of numerical convergence. There are two reasons to suggest that it is indeed physical: The gap persists as we increase the lattice depth, even as the upper solitonic branch contracts to eventually disappear as seen for the $2u$ branch in Fig.~\ref{Fig4_q2_states}.  More significantly, we notice that the phase has different behavior at the tips of the two relevant branches. In Fig.~\ref{Fig4_q2_states}, tip $ml$ of the main branch corresponds to a solution with phase having half the period as that for the solution that corresponds to the tip $2ul$ of the upper $j=2$ branch; both branches are shown in thick red lines for easy comparison; the same pattern is seen in Fig.~\ref{Fig5_q3_states} with their counterparts $ml$ and  $3ul$.

However, the phase issue is a bit more subtle.  In our simulations we determine the solutions along the branches progressively, starting from a specific point in the spectrum, using that solution to find the next adjacent solution and so on.  We find that the density of the eigenstates are insensitive to which initial solution we converge from. On the other hand, the phase can be quite susceptible to the path we take.  We illustrate this in Fig.~\ref{Fig6_phase_var} where we assume lattice periodicty  $q=2$ and strength $V_0=1.3$, as in the center column in Fig.~\ref{Fig4_q2_states}: On the left column, Fig.~\ref{Fig6_phase_var}(a,c,e), we plot the spectrum, density and phase respectively, at specific points marked with filled circles, starting at the tip of the $3d$ soliton branch.  On the right column, Fig.~\ref{Fig6_phase_var}(b,d,f), we plot their counterparts starting at the tip of the $4u$ branch instead.

When we start at the tip of the $3d$ branch, we progress from localized solitonic solutions having density profiles with three nodes. They lose the nodes as we progress towards the main branch; shown in dotted lines of non-uniform modulation in panel (c). On reaching the main branch, we follow it down until we reach the tip labelled $ml$; the corresponding density is shown in thick red line and is a Bloch wave, consistent with Fig.~\ref{Fig4_q2_states}. We then follow the main branch back up all the way to the tip of the $4d$ branch. This path differs from the path of approach; the inset in panel (a) shows a very small gap that exists between the left terminal segment of the $3d$ branch and the main branch in leading down to the tip $ml$. As we follow the main branch up, the densities remain Bloch waves with two density maxima, shown in thin solid lines in panel (c). But as we  approach the $4d$ branch, the two peaks flatten out and each develops a new minimum; so the full density profile develops four minima and eventually four nodes at the tip of the $4d$ branch.

In starting from the $j=4$ branch, for numerical convergence, we have to initiate in the $4u$ branch and progress along it until the solution drops down to the $4d$ branch, and then we continue down to the tip, $ml$. We then converge back up along that branch all the way to the tip of the $4d$ branch and plot the densities and the phases along the path towards $4d$ at exactly the same points as in the previous case, only leaving out the two points on the $3d$ branch, not being present in this path as shown in panel (b).  The densities in panels (c) and (d) for the corresponding points on the spectrum along the two paths of convergence are identical.

However, when we plot the phase in panels (e) and (f),  we find that the phases are completely different for corresponding spectral points along the two different paths of convergence: The right column has half the period of the left for most of the points (only near the tip of the $4d$ branch, the phase on the left path morphs significantly leading up to phase slips at the nodes there). The difference is particularly notable for the point at the bottom tip of the main branch, $ml$, highlighted in thick red lines in Fig.~\ref{Fig6_phase_var}(e) and (f).  This is the upper edge of the gap between the detached $2u$ and main branch in Fig.~\ref{Fig4_q2_states}.

In Sec.~\ref{Sec:Lattice-density} we noted a difference in the phase of the states that correspond to the top and bottom of that gap, evident in the columns corresponding to slightly higher lattice depths $V_0=1.5$ and $1.7$ in Fig.~\ref{Fig4_q2_states}. At those lattice depths, the $3d$ branch has already detached from and extended beyond the main branch, so we could only approach the bottom tip $ml$ of the main branch along the path starting at $4u$ corresponding to the right column panels in Fig.~\ref{Fig6_phase_var}. Now, from the evidence of Fig.~\ref{Fig6_phase_var}  we can conclude that the difference in the phase we observe for above and below the persistent gap between the points $ml$ and $2ul$ is actually an artifact of the different ways we approach those points. This is confirmed by the observation that in the middle column Fig.~\ref{Fig4_q2_states}, for $V_0=1.3$, the phase has the same period for the points above and below that gap. Here the $3d$ branch is still attached to the main branch, and we converged to the point $ml$ along a path starting at the tip of the $3d$ branch corresponding to the left panels in Fig.~\ref{Fig6_phase_var}; the phase period matches that at the bottom edge $2ul$ of the gap which we converged to from the tip of the $2u$ branch.

\section{Interplay of nonlinearity and lattice}

\begin{figure}[t]
    \centering
    \includegraphics[width=1\linewidth]{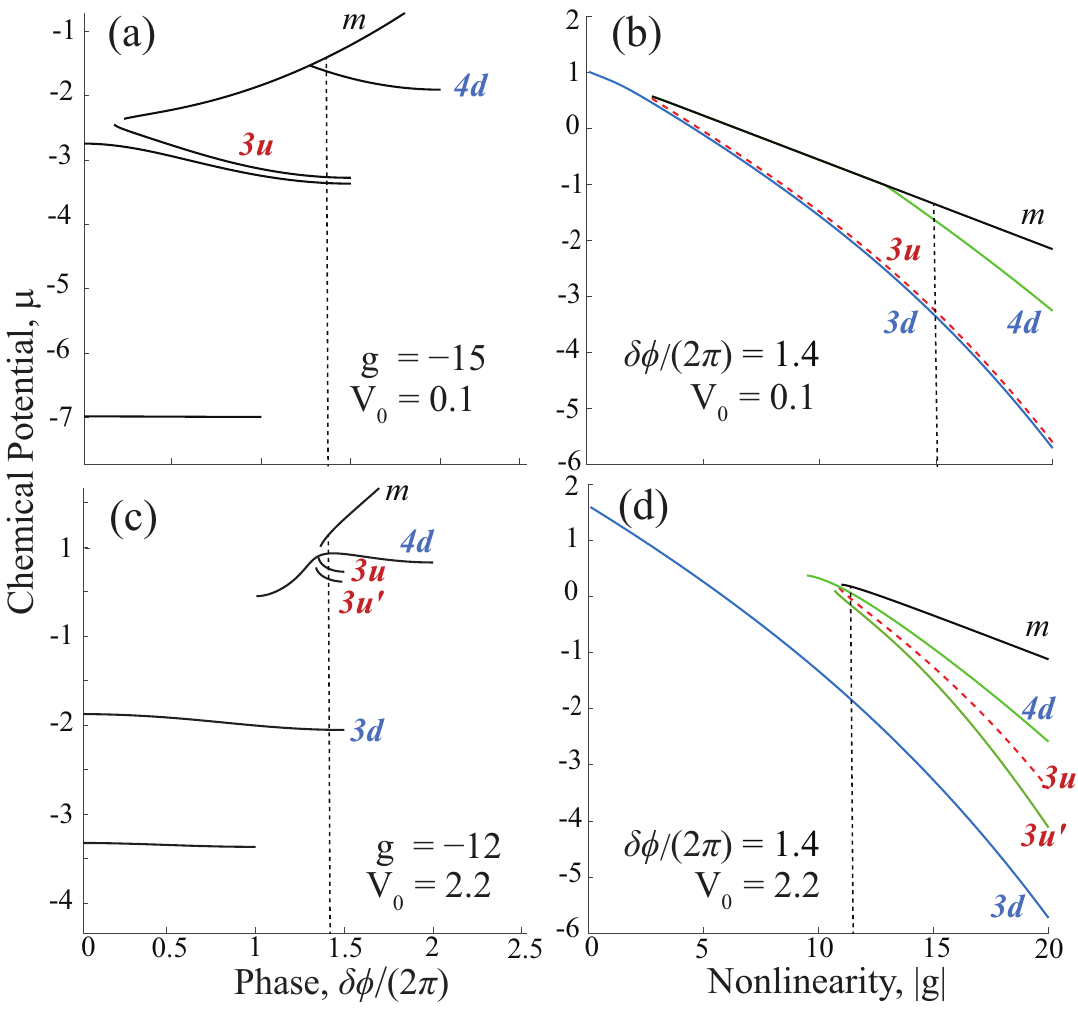}
    \caption{(Color online) Comparison is made for the plots of the chemical potential  as a function of the phase (a,c) at fixed interaction strength, with plots as a function of the interaction strength (b,d) at fixed phase. The lattice periodicity is fixed at $q=3$, with the lattice depth $V_0=0.1$ for the top row and $V_0=2.2$ for the bottom row. The branches are labeled according to our convention in Eq.~(\ref{name-convention}). The vertical dotted lines mark the values in each column that are fixed in the counterpart in the other column.}
    \label{Fig7_vary_g}
\end{figure}

We have so far considered the effects of varying the lattice while keeping the nonlinearity fixed. As we noted in our earlier study of positive nonlinearity \cite{Das-Huang}, the two factors have complementary effects and their interplay defines the spectrum and the states. We re-examine some of the features noted earlier, as we now vary the nonlinearity at fixed lattice depth. In Fig.~\ref{Fig7_vary_g}, we use lattice periodicity $q=3$,  assume a weak lattice $V_0=0.1$ in panels (a,b) and a stronger lattice $V_0=2.2$ in panels (c,d) and plot the spectrum as a function of (i) the bare phase at fixed nonlinearity (left panels) and (ii) the nonlinearity with bare phase fixed at $\phi=1.4$ (right panels).

As we would expect, the chemical potential monotonically decreases as the nonlinearity becomes more negative. Comparing each pair of plots, (a,b) and (c,d), at fixed lattice depth  provides different perspectives. In both cases, the gap between soliton branches of same index, for example $3u$ and $3d$, appears insensitive to the change of  nonlinear strength, with the chemical potential decreasing at about the same rate.  This is in contrast to the mutual gaps between the main branch $m$ and soliton branches associated with different $j$ indices which diverge with stronger interactions. Curiously, in panel (d) we observe the newly emergent $3u'$ branch gradually approach the $3d$ branch as the interaction strengthens.

With both the lattice and the phase fixed in panels (b,d), as a function of the nonlinearity, several of the branches seem to  appear out of the `blue sky' once the nonlinear strength reaches a sufficient strength.  But, seen as a function of phase at fixed nonlinearity in panels (a,c) we can understand most of them as arising from the splitting of soliton branches as the lattice is increased. Together these plots show that the strengths of both the lattice and the nonlinearity work in tandem to create branches at any specific value of the bare phase.

\begin{figure*}[t]
    \centering
    \includegraphics[width=1\linewidth]{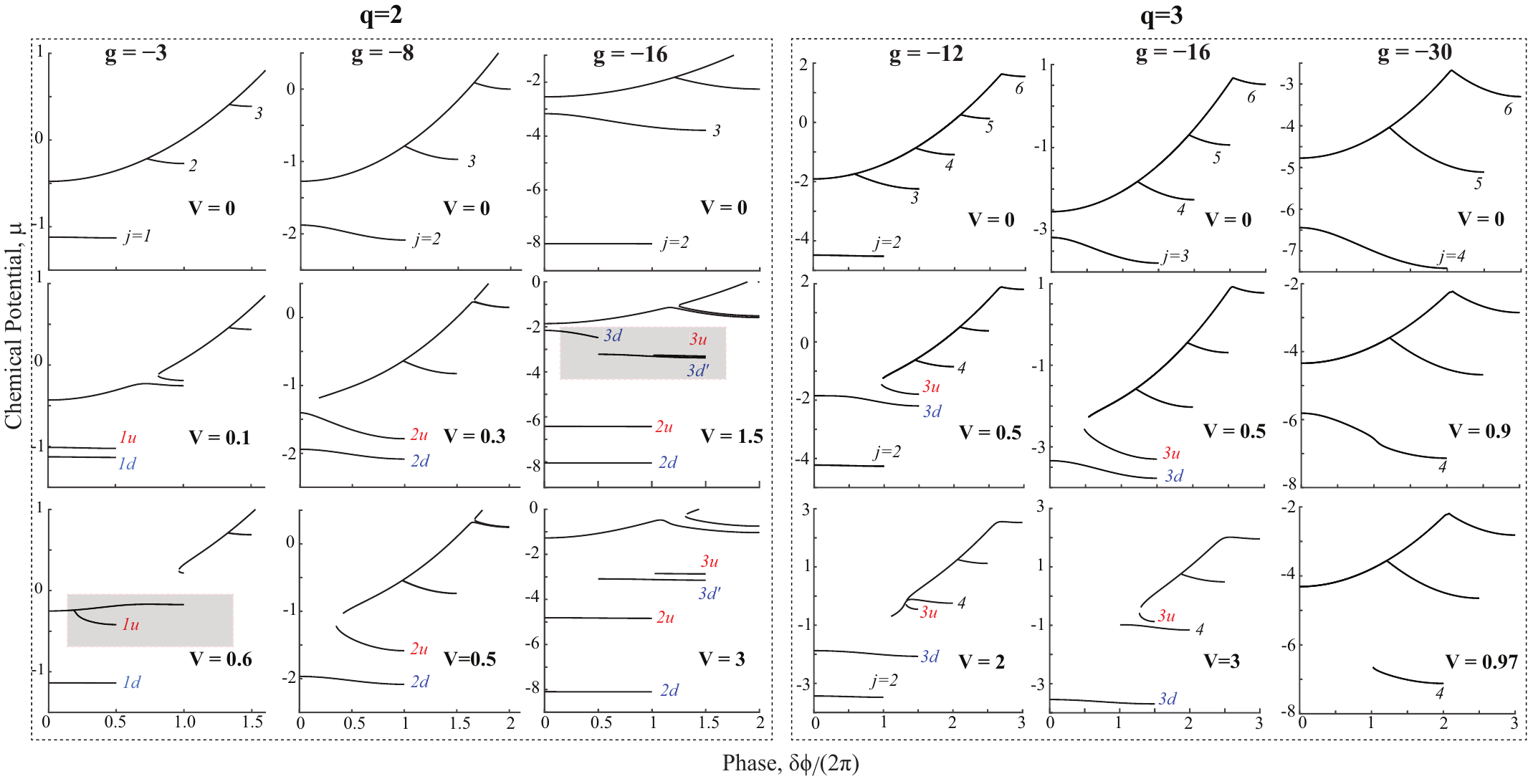}
    \caption{Plots of chemical potential as a function of bare phase with the rows having fixed lattice depth $V_0$ and the columns having fixed nonlinearity $g$, with their values shown in the plots to display their interplay. The left block is for lattice periodicity $q=2$ and the right is for $q=3$, the branches are labeled according to our convention in Eq.~(\ref{name-convention}). The two sections highlighted in gray are examined further in the next figure (Fig.~\ref{Fig9_rejoining}): The lower left shows the rejoining of the $1u$ branch with the main branch, while the other one shows a three-way split of the $j=3$ soliton branch.}
    \label{Fig8_nonlin_lattice_grid}
\end{figure*}

Although in Fig.~\ref{Fig7_vary_g}(a,c) we show snapshots only at two values of the lattice, we tracked the evolution of the spectrum as a function of the bare phase at other lattice depths not shown here. We found that (i) the $4d$ branch detaches from the main branch, it flattens out and eventually spans the phase between $2\pi$ and $4\pi$; (ii) the $3u'$ splits from the $3u$ forming a fork on the right, before separating completely; and (iii) with further increasing of the lattice strength, both of those branches shrink, and eventually disappear, never closing the gap with the main branch.

In Fig.~\ref{Fig7_vary_g}(b,d) the spectrum as a function of the nonlinearity at fixed phase and lattice strength gives a different perspective. At weak nonlinearity, only the main branch remains, underscoring that the soliton branches emerge and extend out from the parabolic dispersion curve as the nonlinearity is strengthened. The left extremes of these branches in panels (b,d) of Fig.~\ref{Fig7_vary_g} mark the threshold nonlinearity at which the corresponding spectral lines, when plotted as a function of $\delta\phi$, reach the value of $\delta\phi/(2\pi)=1.4$ fixed in these panels.

As a different perspective on our discussions in the previous sections, even when the spectrum is plotted as a function of the nonlinearity in Fig.~\ref{Fig7_vary_g}(d), a gap seems to emerge between the main branch and the $3u$ branch. Although the gap is rather small, it is interesting that the $4d$ branch extends into regimes of lower nolinearity beyond the termination of the main branch and $3u$ branch, precisely where those two branches fail to meet. This is consistent with what we observe in the complementary plot in Fig.~\ref{Fig7_vary_g}(c) where that gap is more prominent.

\section{Anomalous spectral features}

The spectrum for negative nonlinearity displays several unusual features that we did not observe with positive nonlinearity. We have noted some of these already, and now we will identify a few others, and also explain the origin of such features. In order to track the interplay of the lattice and the nonlinearity, in Fig.~\ref{Fig8_nonlin_lattice_grid} we plot the spectrum as a function of the bare phase in a grid, where we vary the lattice strength and the nonlinearity in turn while keeping the other fixed, for two different lattice periodicities $q=2$ and $q=3$.

This figure helps explain the curious detachment of certain branches from the vertical axes, accompanied by a similar detachment for the main branch, which we observed earlier in Figs.~\ref{Fig4_q2_states} and \ref{Fig5_q3_states}. We confirm in Fig.~\ref{Fig8_nonlin_lattice_grid} that this occurs only for soliton branches that are commensurate with the lattice period, for example, $j=2$ for $q=2$ and $j=3$ for $q=3$.  More significantly, the relevant commensurate branch needs to be the lowest one still attached to the main branch or be right below the main branch after having detached from it. If the lattice were absent the criterion from Sec.~\ref{Exact_solutions} suggests the range $-(q+1)^2\pi/2<g<-(q-1)^2\pi/2$, but the lattice would shift that. As the magnitude of the nonlinearity is strengthened, branches with higher values of $j$ start `sliding off' and detaching from the main branch. For $q=2$ with no lattice and $V_0=0$, soliton branches $j=1,2$ and $3$  slide off the main branch as the nonlinear strength $g$ gets more attractive from $g=-3$ to $-8$ to $-16$. Likewise, for $q=3$ with no lattice at $V_0=0$, soliton branches $j=2,3$ and $4$ and slide off the main branch as the nonlinear strength $g$ gets more attractive from $g=-12$ to $-16$ to $-30$.  When the lattice is introduced, we observe the main branch detach from the vertical axis along with the $j=q$ upper branch, but only when nonlinearity is in the range to have the two branches be adjacent, as with $g=-8$ column for $q=2$ and $g=-16$ column for $q=-3$. At stronger or weaker nonlinearity this does not occur, for example, at $g=-3$ or $-16$ for $q=2$ and at $g=-12$ or $-30$ for $q=3$.

Comparison among the different columns also clarifies why this happens.  We notice that for commensurate soliton branches which are still attached to the main branch, as the lattice splits the branch, the upper branch forms a hook like structure with the main branch, for example $g=-3$, $V_0=0.1$ for $q=2$ . This is exactly what happens even when the commensurate soliton branch is detached; when the lattice splits it, the upper branch tries to form a similar hook like structure with the main branch. We can see this as the lattice strength is increased from $V_0=0.3$ to $0.5$ for $q=2$ and $g=-8$, and likewise when increasing from $V_0=0$ to $0.5$ for $q=3$ and $g=-16$.  In a variation of this same behavior for $q=3$ and $g=-12$, this occurs when the commensurate branch is the lowest branch still attached to the main branch when the lattice is absent: As the lattice is introduced, at $V_0=0.5$ we see that the main branch has detached from the vertical axis and inches towards the $3u$ branch to form a hook.

One feature persists in these anomalous hook structures, which is that the main branch never seems to join up with the upper of the split commensurate branches,  as we have noted earlier. That remains the case for example with $q=2, g=-8, V_0=0.5$ and for $q=3, g=-12, V_0=0.5$ in Fig.~\ref{Fig8_nonlin_lattice_grid}  This is a point of contrast with such hook structures that appear when the main branch does not detach from the vertical axis as for $q=2, g=-3, V_0=0.1$ in the same figure.

\begin{figure}[t]
    \centering
    \includegraphics[width=1\linewidth]{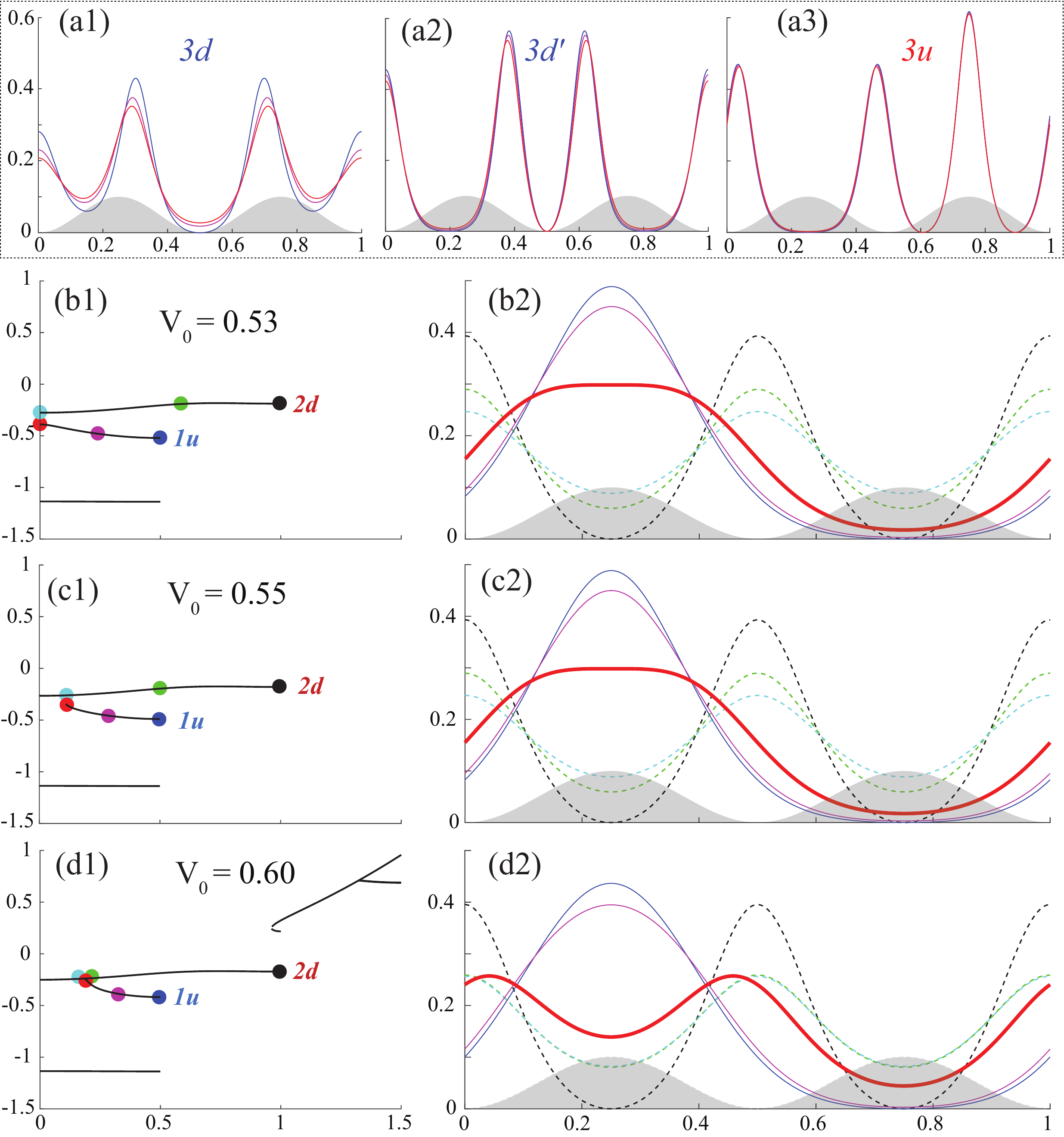}
    \caption{The top row (a1,a2,a3) shows the densities for the three split branches for $j=3$ for $q=2,g=-16,V_0=1.5$, highlighted in Fig.~\ref{Fig8_nonlin_lattice_grid}. The three traces in each panel represent the densities at the tips of each branch and for a point in the middle. The lower panels provide details of the rejoining of the $1u$ branch for $q=2,g=-3,V_0=0.6$, also highlighted in Fig.~\ref{Fig8_nonlin_lattice_grid}. The left panels (b1,c1,d1) show the progression of the $1u$ branch as it detaches from vertical axis and joins the main branch. The right panels (b2,c2,d2) show the evolution of the densities at the points marked on the left panels: The dotted lines mark Bloch waves corresponding to the $2d$ branch. The solid lines correspond to the $1u$ branch, with the thick red one marking the left tip of that branch.}
    \label{Fig9_rejoining}
\end{figure}

In Fig.~\ref{Fig8_nonlin_lattice_grid} we use gray shading to highlight two other anomalous spectral features: First, for $q=2$ and $g=-3$, increasing the lattice strength from  $V_0=0$ to $V_0=0.6$ causes the $j=1$ branch to split; the gap between the branches widen, until the upper branch, $1u$ actually rejoins the evolved main branch. Secondly, for $q=2, g=-16$ and $V_0=1.5$, we find the $j=3$ branch develops multiple splittings that we label $3d, 3d'$ and $3u$. The $3d$ branch disappears as the lattice is strengthened to $V_0=3$, and the two branches $3u$ and $3d'$ remain detached from both the main branch and the vertical axis. Our labelling of these branches is based their symmetry properties as defined earlier in Sec.~\ref{Sec:Lattice-density}. This is illustrated in Fig.~\ref{Fig9_rejoining}(a1,a2), where the densities for the $3d$ and $3d'$ branches each have one minimum that coincides with a lattice minimum, characteristic of the energetically lower of a split branch. In contrast, that feature is absent for the density of the $3u$ branch shown in Fig.~\ref{Fig9_rejoining}(a3), which additionally does not have any of its density maxima aligned with a lattice minimum, characteristic of energetically higher of a split branch.

We can surmise that the two branches $3d$ and $3d'$ split at what emerged as an inflection point at a lower lattice strength.  We can see something similar occurring for the $j=4$ branch in Fig.~\ref{Fig8_nonlin_lattice_grid} for $q=3$ at lattice depth of $V_0=0.9$. At higher lattice depth, $V_0=0.97$, it is quite likely that there are two separate branches that have split at that inflection point, but we could only converge to the right branch as shown.

In the lower panels (b-d) of Fig.~\ref{Fig9_rejoining}, we examine in further detail the rejoining of the $1u$ branch highlighted in the lowest left panel of Fig.~\ref{Fig8_nonlin_lattice_grid}. We plot the densities of a few points on the main branch as well as along the $1u$ branch as it rejoins the main branch.  We label the main branch as $2d$ in this figure, since Fig.~\ref{Fig8_nonlin_lattice_grid} shows that at these lattice depths, the right termination of the main branch is the $2d$ branch. Before joining, at $V_0=0.53$, the densities (shown in solid lines) along the $1u$ branch have one minimum, as we would expect with $j=1$, while the densities along the main branch (shown in dotted lines) are characteristic of Bloch waves with two maxima, as it should be with $q=2$. However we note at the left end of the $1u$ branch still on the vertical axis, the density, shown in thick red line in panel (b2), is noticeably flattened. This remains so as the branch detaches from the vertical axis at $V_0=0.55$. But, when it joins the main branch at $V_0=0.60$, the flattened peak morphs to develop a new minimum, heralding that it is now on the main branch. But this point of joining is not a Bloch wave and is markedly nonuniform in its modulation, having characteristics of both  branches that meet here.

\begin{figure*}[t]
\centering
    \includegraphics[width=1\linewidth]{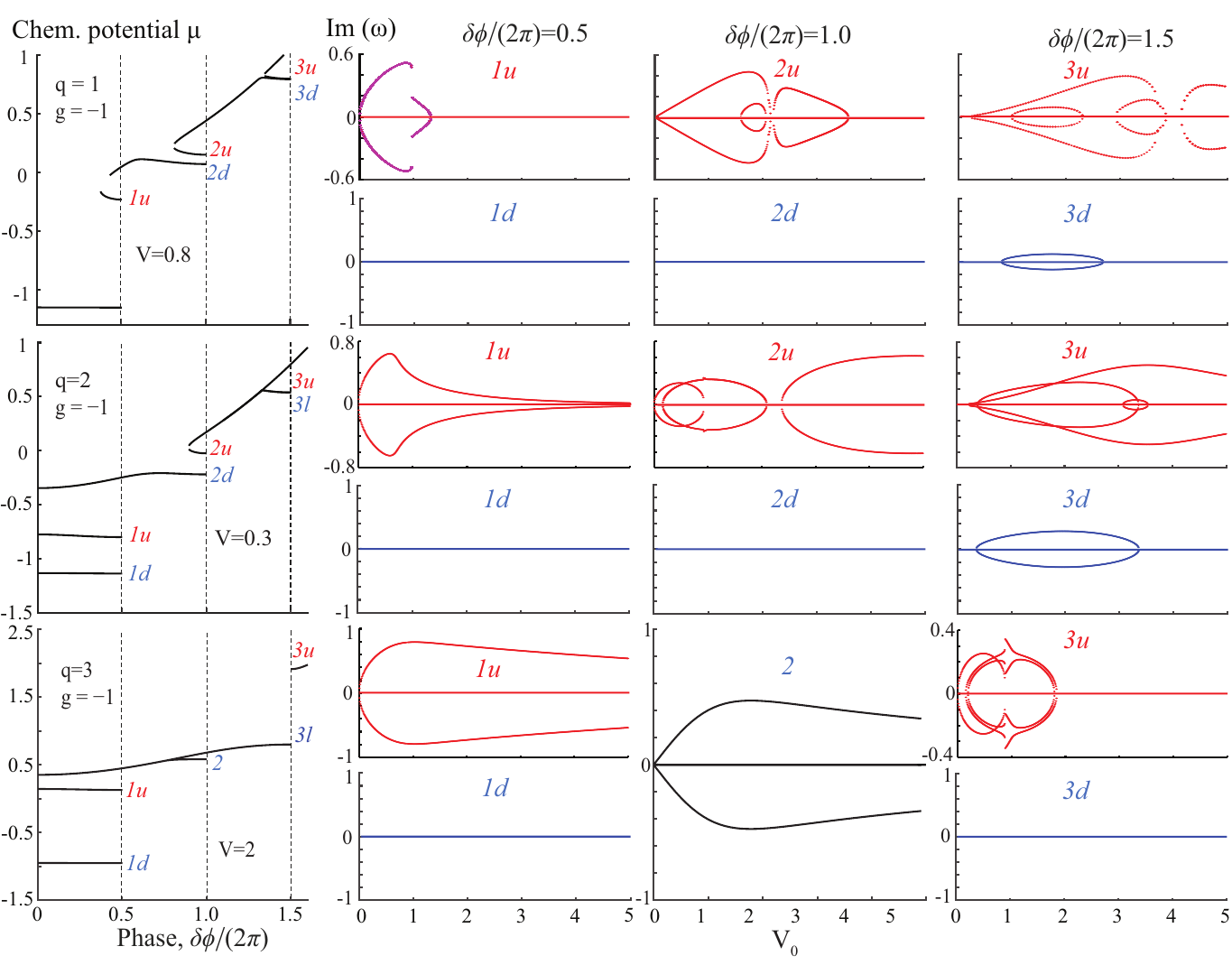}
    \caption{The measure of instability of the eigenstates as a function of the strength ($\pm V_0$) is gauged by the imaginary part of the normal modes ${\rm Im}(\omega)$ of Bogoliubov fluctuations. The three rows have lattice periodicity $q=1,2,$ and $3$ respectively,  for a low nonlinear strength $g=-1$.  The leftmost panels show a \emph{representative} spectrum at a specific potential indicated, with vertical lines indicating the bare phase values, $\delta\phi/(2\pi)=0.5,1,$ and $1.5$, for which  the modes are plotted in the other three columns. The upper and lower branches are plotted separately, except for the $j=2$ branch for $q=3$ in bottom row, since that branch was not observed to split. Non-vanishing ${\rm Im}(\omega)$ marks instability.}
    \label{Fig10_stability}
\end{figure*}

\section{Stability of States}
We will now examine the dynamical stability properties of the solutions we found, by considering small perturbation around the mean field stationary states:
\bn \psi(\theta,t)=\psi_0(\theta)+\delta u e^{-i\mu t} e^{-i\omega t}+\delta v^* e^{-i\mu t} e^{i\omega^* t}\en
and then solve the Bogoliubov equations \cite{RMP-Sringari-1999} for the normal modes of the fluctuations.
\bn (H_0+2g|\psi_0|^2-\mu)\delta u+ g\psi_0^2\delta v=\omega \delta u\n\\
-(H_0+2g|\psi_0|^2-\mu)\delta v+ g(\psi^*_0)^2\delta u=\omega \delta u\en
If the angular frequencies $\omega$ of the normal modes have  imaginary components and if ${\rm Im}(\omega)>0$ then the fluctuations would grow exponentially, indicating dynamical instability.

Figure~\ref{Fig10_stability} shows the stability properties for a relatively weak nonlinear strength of $g=-1$, for three different lattice periodicities $q=1,2,$ and $3$, and the lowest soliton branches $j=1,2,$ and $3$ along the respective rows. The left column shows the spectrum for each periodicity at a fixed lattice depth.  The remaining three columns plot the imaginary part ${\rm Im}(\omega)$ of the eigenmodes of the Bogoliubov fluctuations as the lattice depth is varied.  Dotted lines on the spectrum mark the bare phase values $\delta\phi/(2\pi)=0.5, 1, 1.5$ for which the modes are plotted; they correspond to the tips of branches, the solutions with nodes. The vast majority of the modes,  for every case shown are purely real with ${\rm Im}(\omega)=0$ which is represented by the horizontal line present in all the plots. In most of the plots, there are a few modes with non-zero ${\rm Im}(\omega)$, always the lowest one or two modes.

With no lattice present, all of the solutions displayed are stable with all the modes being real, consistent with conclusions in previous studies for repulsive interaction \cite{carr09,Das-Huang}.  As the lattice is strengthened, the lower branches, $j=1d,2d$ and $3d$ remain stable at least for weak lattices. The only exceptions are the $3d$ branches for $q=1,2$ and the $j=2$ branch for $q=3$, which remains unsplit as far as we could tell. Curiously, in the former case the instability appears only for a certain range of lattice depth. This is reminiscent of periodic patterns we observed for nonlinear stationary solutions with a barrier, caused by the variation of the density at points of transition of the potential \cite{Brattley-Das-PRA}. We tested the stability up to lattice depths of $V_0=20$ and the lower branches remained stable.

In contrast, the upper branches are unstable as soon as the lattice is turned on, characterized by at least some nonzero ${\rm Im}(\omega)$.  Their pattern of variation with the lattice depth is diverse. A few key features are apparent: As the lattice gets very deep the instability seems to decline; there also appear to be some islands of stability at certain lattice depths, analogous to the regimes of instability we observe for the lower branches. We do not yet have an explanation for why those occur at those specific values, and this deserves further investigation.

Our main observation regarding the dynamical stability behavior is that the behavior for negative nonlinearity is the opposite of that for positive nonlinearity \cite{Das-Huang}, where we observed that the upper branches were generally stable in the presence of the lattice. In fact, the behavior was qualitatively identical as described above, except that the stability of the upper and lower branches are switched.  This underscores that the sign of the nonlinearity, as in whether interaction is attractive or repulsive, directly impacts the stability properties of the soliton branches.

\section{Conclusions and Outlook}

Since the calculations with lattice are obtained numerically, we comment briefly on some of the challenges. We find the solutions using Newton's method in Fourier space, where we use a finite number of basis states, which we have described in our earlier paper \cite{Das-Huang}. Stronger nonlinearity requires a larger basis, and curiously we found that negative nonlinearity requires more basis states than positive nonlinearity of the same magnitude. We first find the solutions with no angular velocity $\Omega=0$ and then increment $\Omega$ gradually to build the full spectrum. This has generally worked very well, but when there are strong inflections or a jump in the branch, the method can fail to find the continuation of a branch. Also, it is possible to miss branches that may not terminate at a node. We tried to account for these limitations, but despite our best efforts some spectral features or lines may be missing, particularly at stronger nonlinearity which are computationally more demanding.

Our examination of the negative nonlinearity induced by attractive interaction complements our prior work with positive nonlinearity \cite{Das-Huang} and completes our survey of the landscape of stationary states of ultracold atoms trapped in quasi 1D ring shaped lattices. In the absence of an azimuthal lattice, the analytical solutions display a convenient mapping as the sign of the nonlinearity is changed. When the lattice is introduced, the solutions are found to present symmetry and stability properties that are in many respects the opposite between positive and negative nonlinearities, but not completely. While certain features like the splitting of soliton branches by the lattice and general impact of an angular velocity remain unaffected by the change of sign of the nonlinearity, there are various other features that are specific to attractive nonlinearity.  The most prominent of those is the progressive detachment and `falling off' of the soliton branches from the main parabolic one, leading to new ground states.  When a solitonic branch with index commensurate with the lattice periodicity is the lowest one still attached or the highest one detached and energetically right below the main parabolic branch, these branches are found to separate from the vertical axis and tend to form a hook like spectral pattern. Among other spectral features observed, soliton branches can undergo multiple splits and even rejoin after detachment.

In the case of attempted rejoining of detached commensurate branches with the main parabolic branch, we found certain persistent gaps.  The phase of the states at the edge of these gaps is found to be sensitive to the path of convergence taken to approach those particular spectral points.  This can be directly translated to adiabatic evolution along different paths by dynamically evolving the system parameters, like angular velocity and lattice depth and interaction strength. Examining the path dependence of the phase under such variations is part of our ongoing work.

The symmetry properties are not quite mutually opposite between positive and negative nonlinearities, but they are distinct. Likewise the stability features are generally complementary with respect to the behavior of the upper and lower branches in a lattice-induced splitting. The localization properties of the states are the same regardless of the sign of the nonlinearity, with soliton branches that are incommensurate with the lattice periodicity having solutions that tend to localize on introducing the lattice.

Together with our study of the case of repulsive nonlinearity \cite{Das-Huang} this paper will provide a firm theoretical basis for exploring the dynamics of ultracold atoms in ring shaped traps particularly in the substantially more interesting case of having an azimuthal lattice. This system offers the possibility of examining numerous interesting phenomena in a system that brings together elements of quantum mechanics, topology, nonlinear dynamics and mesoscopic condensed matter physics. The intrinsic description in terms of a NLSE also broadens the possibility of applications to fiber optical systems as well.  We are currently working on applying our results to understand the dynamics of coherent median in ring-shaped lattices, and we hope to motivate experiment that will explore this very rich system.

\begin{acknowledgments} We gratefully acknowledge the support of the NSF under Grant No. PHY-2011767. We thank Prof. Dominik Schneble, Hongyi Huang and Allison Brattley for valuable discussions. \end{acknowledgments}

%\bibliography{ring_neg_nonlin}

\end{document}